\documentclass[a4paper,fleqn]{cas-sc}
\usepackage[utf8]{inputenc}
\usepackage[T1]{fontenc}

\usepackage[numbers]{natbib}

\usepackage{siunitx}
\usepackage{caption}
\usepackage{subcaption}

\def\tsc#1{\csdef{#1}{\textsc{\lowercase{#1}}\xspace}}
\tsc{WGM}
\tsc{QE}

\begin{document}
\let\WriteBookmarks\relax
\def\floatpagepagefraction{1}
\def\textpagefraction{.001}

\shorttitle{Performance of the prototype Silicon Tracking System of the CBM experiment tested with heavy-ion beams at SIS18}    

\shortauthors{}  

\title [mode = title]{Performance of the prototype Silicon Tracking System of the CBM experiment tested with heavy-ion beams at SIS18}  



\author[1]{A.~Agarwal}
\author[1]{Z.~Ahammed}
\author[2]{N.~Ahmad}
\author[3]{L.J.~Ahrens}
\author[4]{M.~Al-Turany}
\author[2]{N.~Alam}
\author[5]{J.~Andary}
\author[6]{A.~Andronic}
\author[5]{H.~Appelsh\"{a}user}
\author[5]{B.~Arnoldi-Meadows}
% ...
\affiliation[1]{organization={Variable Energy Cyclotron Centre (VECC), Kolkata, India}}
\affiliation[2]{organization={Department of Physics, Aligarh Muslim University, Aligarh, India}}
\affiliation[3]{organization={Justus-Liebig-Universit\"{a}t Gie{\ss}en, Gie{\ss}en, Germany}}
\affiliation[4]{organization={GSI Helmholtzzentrum f\"{u}r Schwerionenforschung GmbH (GSI), Darmstadt, Germany}}
\affiliation[5]{organization={Institut f\"{u}r Kernphysik, Goethe-Universit\"{a}t Frankfurt, Frankfurt, Germany}}
\affiliation[6]{organization={Institut f\"{u}r Kernphysik, Universit\"{a}t M\"{u}nster, M\"{u}nster, Germany}}
% ...

%
















\begin{abstract}
    The Compressed Baryonic Matter (CBM) experiment at the future Facility for Antiproton and Ion Research (FAIR) is a heavy-ion experiment designed to study nuclear matter at the highest baryonic density. For high-statistics measurements of rare probes, event rates of up to 10 MHz are targeted. The experiment, therefore, requires fast and radiation-hard detectors, self-triggered detector front-ends, free-streaming readout architecture, and online event reconstruction.
     
    The Silicon Tracking System (STS) is the main tracking detector of CBM, designed to reconstruct the trajectories of charged particles with efficiency larger than 95\%, a momentum resolution better than 2\% for particle momenta larger than 1 GeV/c inside a \qty{1}{Tm} magnetic field, and to identify complex decay topologies. It comprises 876 double-sided silicon strip modules arranged in 8 tracking stations.
    
    A prototype of this detector, consisting of 12 modules arranged in three tracking stations, is installed in the mini-CBM demonstrator. This experimental setup is a small-scale precursor to the full CBM detector, composed of sub-units of all major CBM systems installed on the SIS18 beamline. 
    In various beam campaigns taken between 2021 and 2024, heavy ion collisions at 1--\qty{2}{AGeV} with an average collision rate of \qty{500}{kHz} have been measured.
    
    This allows for the evaluation of the operational performance of the STS detector, including time and position resolution, hit reconstruction efficiency, charge distribution, signal-to-noise ratio, and its potential for track and vertex reconstruction. 
\end{abstract}


\maketitle

\section{Introduction} \label{introduction}
    \subsection{The CBM experiment} \label{sec:cbm}
        The Compressed Baryonic Matter (CBM) is a fixed target heavy-ion experiment at the Facility for Antiproton and Ion Research (FAIR) under construction in Darmstadt, Germany. It will investigate the phase diagram of strongly interacting matter in the region of high net-baryon densities and moderate temperatures \cite{CbmPhyProg2017}. For high-statistics measurements of rare probes, event rates of up to \qty{10}{MHz} are targeted. To meet these demands, the CBM experiment uses fast and radiation-hard detectors, self-triggered detector front-ends, and a free-streaming readout architecture.
    
    \subsection{The STS detector} \label{sec:cbm_sts}
        The Silicon Tracking System (STS) is the core detector for tracking and momentum determination. It consists of 876 double-sided silicon micro-strip detectors arranged in eight tracking stations positioned between \qty{30}{cm} and \qty{100}{cm} downstream of the target inside a \qty{1}{Tm} magnetic dipole field \cite{StsTDR2013}. The primary purpose of the STS is the reconstruction of the trajectories of up to $\sim$$10^3$ charged particles per beam-target collision event. This should be achieved with an efficiency larger than 95\% and momentum resolution better than 2\% for primary particles. It also enables the reconstruction of complex event topologies, such as the weak decays of strange or charmed hadrons and hypernuclei. To accomplish these goals at the projected interaction rate, the STS requires a position resolution better than \qty{30}{\um} in the bending plane, a good time resolution ($<\qty{10}{ns}$), low per-channel dead-time and a material budget within 0.3\%--1.4\%~$X_0$ per tracking station \cite{StsTDR2013}.
    
    \subsection{The STS module} \label{sec:sts_module}
        The functional building block of the STS is the module \cite{RodriguezRodriguez2024}. It comprises a Double-Sided Double-Metal (DSDM) silicon micro-strip sensor and associated readout electronics. The sensor, produced by Hamamatsu Photonics, K.K., Japan \cite{HamamatsuPhotonics}, are \qty{320}{\um} thick and \qty{62}{mm} wide and have four different lengths: \qty{22}{mm}, \qty{42}{mm}, \qty{62}{mm} and \qty{124}{mm}. Each sensor has 1024 strips per side, spaced at \qty{58}{\um} pitch. The strips on the p-doped side (p-side) are inclined by \qty{7.5}{^\circ} with respect to the n-side, thus providing 2D tracking information with a single sensor. Each sensor side is read out by a front-end board (FEB), equipped with a total of eight custom-designed STS/MUCH-XYTER ASICs (SMX) \cite{KASINSKI2018225}. Each of these is connected to 128 strips of the sensor via a pair of custom ultra-lightweight aluminium-polyimide microcables.
        
        The SMX ASIC utilizes parallel signal processing to generate signals independently for each channel, featuring a charge-sensitive amplifier (CSA) that feeds into a dual signal path with two shaper amplifiers: a fast shaper for timestamp generation and a slow shaper for amplitude measurement. Each of the 128 analog channels incorporates a 5-bit continuous flash analog-to-digital converter (ADC) to measure signal amplitude, along with a fast leading-edge discriminator to determine the signal's generation time in a 14-bit timestamp message with a resolution of \qty{3.125}{ns}. The chip is based on a streaming data processing architecture that allows operation without data loss in the digital readout path with an average load of up to $2.8\times10^5$ signals per channel per second. The readout can be performed with up to five synchronous serial data links per ASIC. A new signal arriving while processing a prior signal is lost, and an Event-Missed flag is activated \cite{KASINSKI2018225}.
    
    \subsection{The mCBM setup} \label{ssec:mcbm_setup}
        Mini-CBM (mCBM) is a demonstrator test setup consisting of prototype or pre-series components of all major CBM systems installed at the SIS18 synchrotron of the GSI Helmholtz Center, constructed to study, commission, and test the complex interplay of the different detector systems, the readout electronic chain with free-streaming data acquisition and fast online event reconstruction and selection \cite{mCBM_SIS18_Proposal}. The experiment has been operational at SIS18 since 2018 and measures heavy ion collisions at 1--\qty{2}{AGeV}. The mSTS setup was active in various beam test campaigns between 2021 and 2024.
        
        This paper presents a detailed study of the operational performance of the prototype STS system measured in beam experiments with the mCBM setup. 

            \begin{figure}[htbp] 
            \centering
            \includegraphics[width=0.8\linewidth]{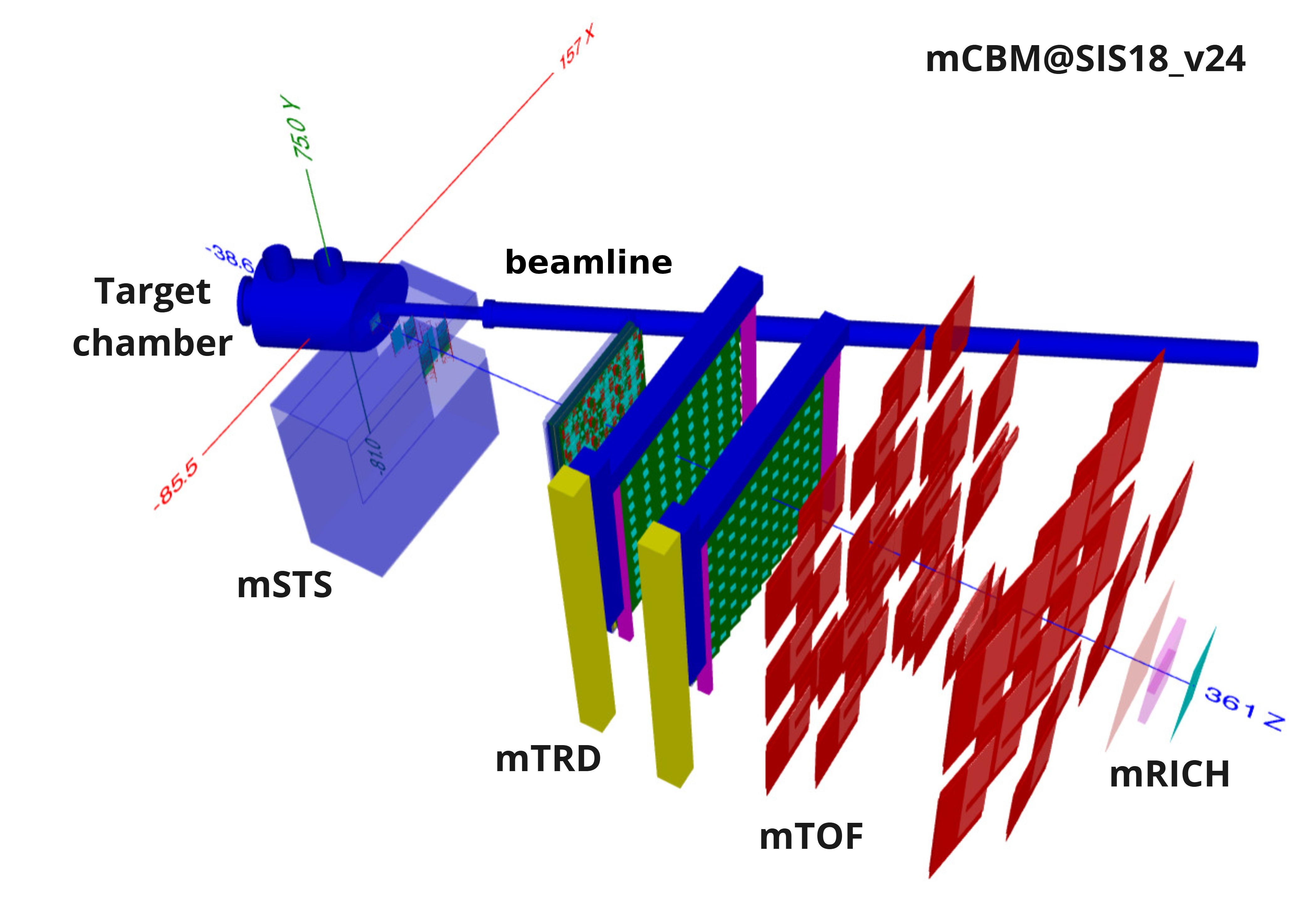}
            \caption{Schematic view of the mCBM setup in the 2024 beam campaign.}
            \label{fig:mcbm_setup_2024}
        \end{figure}

\section{Experimental conditions} \label{sec:exp_conditions}
    \subsection{Experimental setup} \label{ssec:sts_setup}
        Figure~\ref{fig:mcbm_setup_2024} presents the schematic of the experimental configuration utilized in this analysis. The primary axis of the mCBM setup (z-axis) is positioned at an angle of $25^{\circ}$ relative to the beam axis. With a right-handed coordinate system, the positive y-axis is oriented upwards, and the positive x-axis is directed to the left when looking downstream.
                
        It includes the Beam Monitoring detector prototype (BMon), which serves to measure the start time of the collision. BMon, not shown in Fig.~\ref{fig:mcbm_setup_2024}, but detailed in a zoom-in view of the target chamber later in Fig.~\ref{fig:mcbm_target}, consists of a $1\times\qty{1}{cm^2}$ polycrystalline diamond detector placed \qty{20}{cm} upstream of the target, denoted as T0. In 2024, a second diamond, denoted as B1, was mounted \qty{45}{cm} upstream of the target. Three layers of the transition radiation detector (TRD)~\cite{TrdTDR2018} are mounted downstream of the STS, followed by up to four layers of Time-of-Flight (TOF) counters~\cite{TofTDR2014}. A Ring-Imaging Cherenkov (RICH) detector is installed downstream~\cite{RichTDR2013}.
        
    \subsection{The STS system}
\begin{figure}[h] 
            \centering
                \includegraphics[width=.7\linewidth]{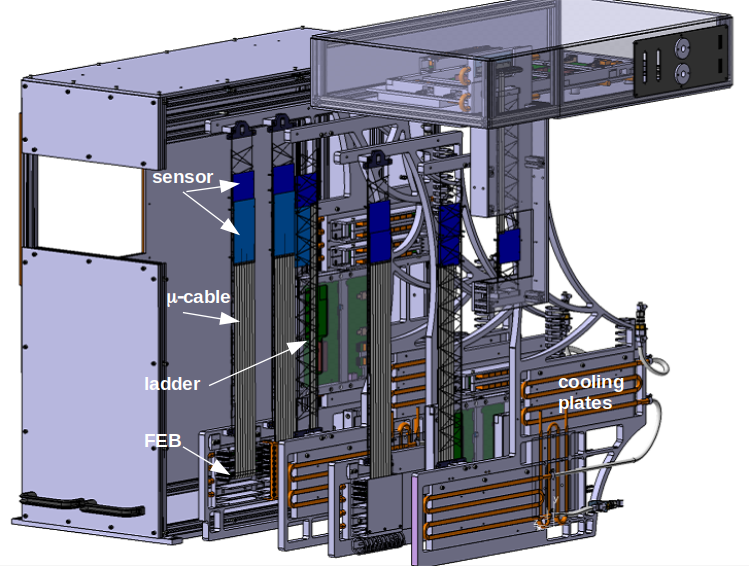} 
            \caption{CAD drawing of the mSTS setup shifted outside the enclosure for illustration. The sensors are depicted in blue.}
            \label{fig:sts_setup_2024}
        \end{figure}

        The STS setup installed at the mCBM experiment is shown in detail in Fig.~\ref{fig:sts_setup_2024}. It consists of three tracking stations of an active area of $6\times\qty{6}{cm^2}$, $12\times\qty{12}{cm^2}$, and $18\times\qty{18}{cm^2}$, positioned respectively around \qty{16}{cm}, \qty{33}{cm}, \qty{47}{cm} from the target. The stations are constructed from 12 detector modules of two different sensor sizes, $62\times\qty{62}{mm^2}$ and $62\times\qty{124}{mm^2}$. Modules were integrated into the setup in multiple stages. All modules employ series sensors and microcables, along with close-to-final or series-production SMX ASICs. The earliest modules still rely on prototype FEBs, powering, and bias voltage filtering. The ASICs in the two downstream stations use only 1 uplink, while the ASICs in the first upstream station use all 5 available uplinks. Seven Common Read-Out Boards (C-ROB) are used for the readout, and five Power Boards (POB) provide the supply voltages for the FEBs. The modules are mounted on carbon-fiber ladders, which are affixed to aluminum support structures (C-frames) where the front-end electronic boards are placed.
        
    \subsection{Commissioning and operation} \label{ssec:operation}
        Before installation in the mCBM cave, an energy calibration was carried out to adjust the dynamic range of the per-channel ADCs. This was done using a calibrated internal pulse generator with gradually increasing amplitudes in the readout chip to record the response of each discriminator. This provides the effective threshold and gain for each detector channel. The derivative of the discriminator response in the S-curve scan provides a way to estimate the baseline noise level. An Equivalent Noise Charge (ENC) of \qty{1000}{e}, below the targeted system noise, was measured \cite{RodriguezRodriguez2024}. The modules were operated with relatively low thresholds, around 3--\qty{4}{\sigma} RMS amplitude of the noise, i.e., 3000--\qty{4000}{e}. The typical dark rate observed was around \qty{0.5}{kHit/s/channel} and up to \qty{1.4}{kHit/s/channel} in the worst case, consistent or better than the expectations \cite{StsTDR2013}.
    
    \subsection{Data sets} \label{ssec:data_set}
        Different beam campaigns have been performed with the mCBM setup since 2021 with heavy-ion collisions in fixed-target configuration, typically using mid-size ion species (O, Ni) at 1--\qty{2}{AGeV} and a moderate beam intensity of about \qty{500}{kHz}. Additionally, tests with larger colliding systems (Au+Au, U+Au) at higher interaction rates were conducted. 
        
        Most of the data presented here were collected in 2024 in Ni+Ni collisions with a beam kinetic energy of \qty{1.93}{AGeV}. The beam intensity was about $4\times10^{7}$ ions per spill, and a spill length of about \qty{8}{s}, followed by a spill-break of about \qty{2.5}{s}. The beam profile had a width of \qty{1.5}{mm}. The Ni-target employed in the experiment measures \qty{15}{mm} $\times$ \qty{35}{mm}. It is mounted in a thin Al frame with a cutout of \qty{31}{mm} in diameter. The target thickness of \qty{4}{mm} corresponds to 10\% interaction probability, resulting in a collision rate of about \qty{500}{kHz}.
        
    \subsection{Data analysis} \label{ssec:data_analysis}
        In line with the free-streaming readout architecture, the data is collected and recorded in Time Slice Archive (TSA) files. A time slice contains the full data recorded by the DAQ~\cite{CuvelandDAQ2022} in a defined time frame from every detector. In the offline analysis, the TSA files are first decoded into strip-raw signals for the STS, which contain the module and channel information of the activated strip, as well as the signal's time and amplitude. Clusters are reconstructed by correlating signals from neighboring activated strips~\cite{FrieseClusterFinding2019}, within a short time window. The amplitudes of signals from individual strips of a cluster are summed up. Time information is the average of the individual signal's time. The center of gravity of a cluster is taken as an estimate for the crossing point of a particle in the sensitive volume of the detector \cite{Malygina_2017}. Hits, i.e., space points, are finally derived by correlating sensor p- and n-side clusters. Events are defined using the signal from a seed detector (BMon in this case) as a time reference and including all signals from the other detectors within a time window that depends on the detector's time resolution. In this prototype beam campaign, a conservative window of $\pm$\qty{60}{ns} was used for the STS. Only a single BMon hit is required to prevent event pile-up. A minimum of eight TOF signals is required to remove empty events triggered by sporadic noise counts and enhance the sample that contains at least one track. Events with more than 100 STS signals are removed to suppress events caused by massive pickup noise. 
    
\section{Results} \label{sec:results}
    \subsection{Throughput of the readout electronics} \label{ssec:readout}
        The throughput of the readout electronics was studied in beam intensity scans with U+Au collisions. In these beam campaigns, only the modules built with one data uplink per ASIC were installed. The maximum data rate reached in measurement was \qty{56}{kHit/s/channel}. This is consistent with the expected bandwidth saturation limit of \qty{9.41}{kHit/s/link} (corresponding to \qty{73}{kHit/s/channel}), measured in the laboratory \cite{KASINSKI2018225}, but limited in beamtime by the microscopic spill structures of the beam.
        The throughput of the module readout electronics can be increased by a factor of five by using 5 links per ASIC, thereby reaching a bandwidth of \qty{370}{kHit/s/channel} to cope with the high-intensity heavy-ion beams expected at CBM.
        
        While reaching the maximum throughput, a further increase of beam intensity leads to an observed increase of the Event-Missed flag. Typical values of the data loss rate for the beam intensity used in the performance studies with Ni+Ni data at 500 kHz are well below 1\%. 
        
        \begin{figure}[!h] 
            \centering
            \includegraphics[width=.47\textwidth]{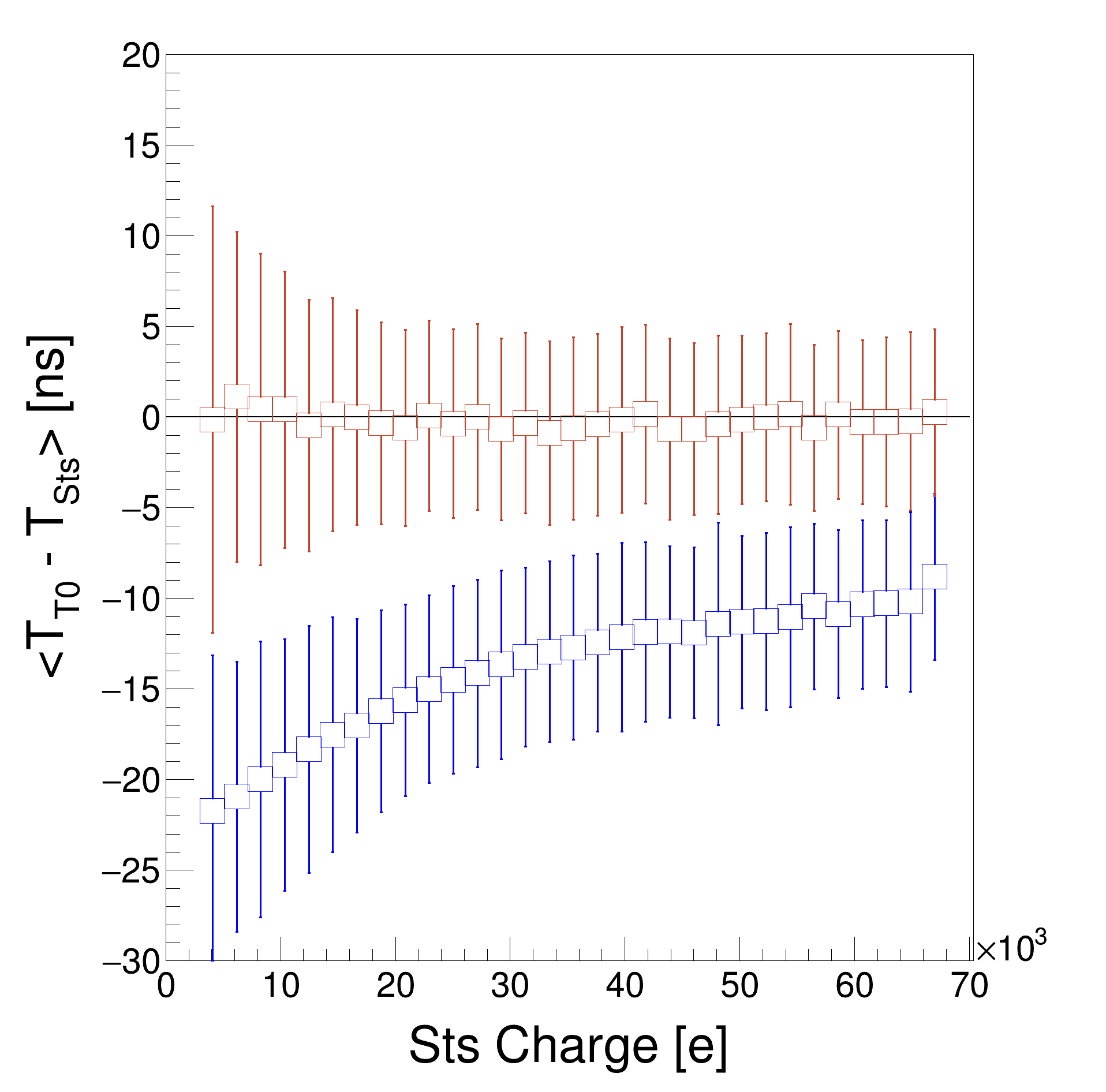}\quad
            \includegraphics[width=.47\textwidth]{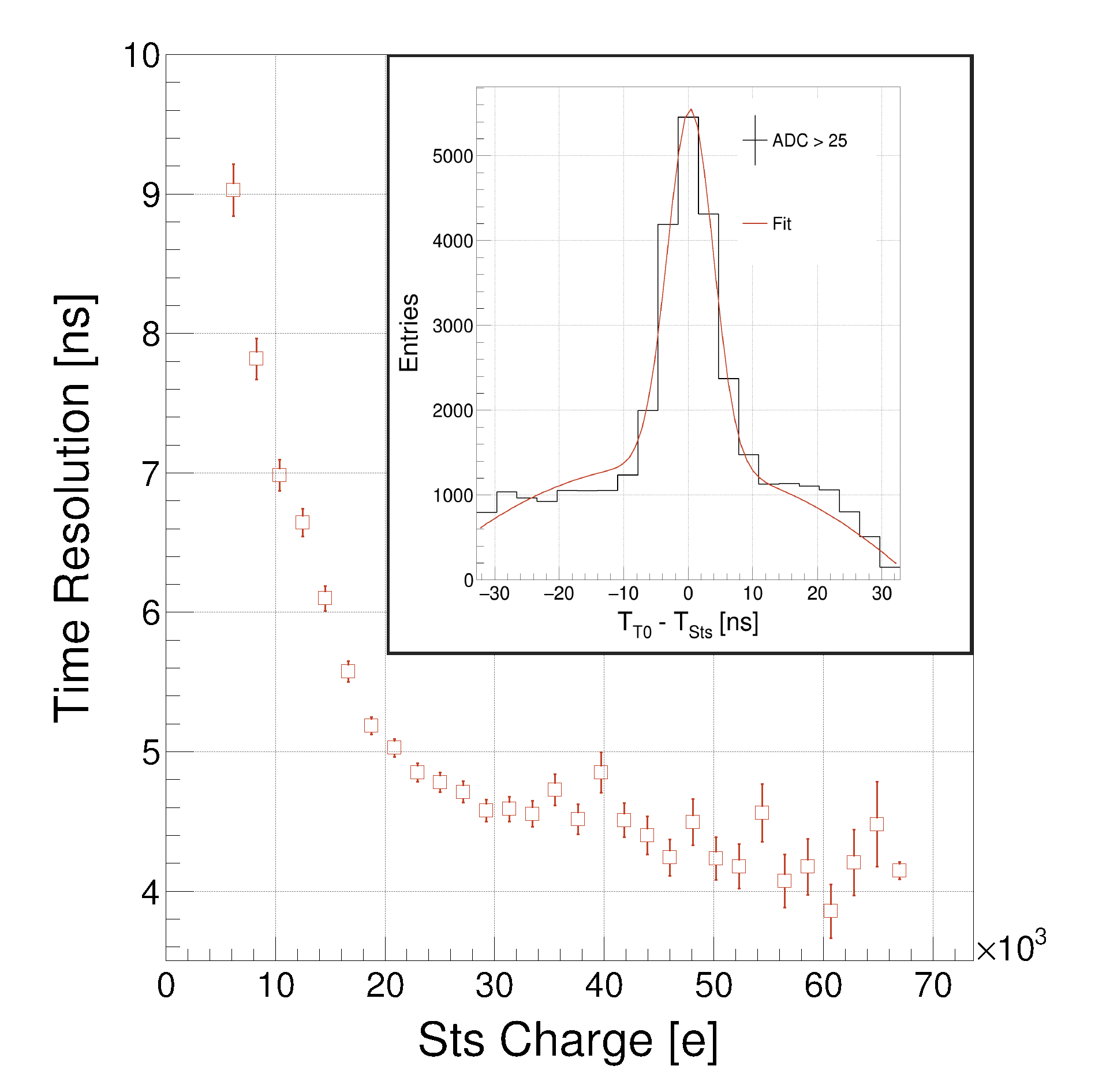}\\
            \caption{Top: Time difference of raw strip signals of the STS and signal from T0 detector, before (blue) and after (red) time calibration. The points indicate the mean value of the time difference, while the bar stands for the width ($\sigma$) of the distribution. Bottom: Time resolution of the STS as a function of the signal amplitude. The error bars reflect the fit uncertainty. The inset shows the time difference between STS and T0 signals for large STS signal amplitude, $>$ 25 ADC.}
            \label{fig:time}
        \end{figure}
    
    \subsection{Time resolution} \label{time_resolution}
        The detector raw signals are synchronized to ensure proper functionality in a free-streaming operation. Individual subsystem time-offset corrections calculated from a common reference are applied. The T0 detector, part of the BMon system, is used as a reference. The STS time calibration accounts for the time offset and the timing's dependence on the signal amplitude (time walk). It is performed for every ASIC and discriminator.
        
        Figure~\ref{fig:time}, left part, shows the average time difference between STS and T0 signals as a function of the STS signal amplitude before and after calibration. The points indicate the mean value of the time difference, while the bar stands for the width of the distribution obtained via a Gaussian fit. After the calibration is applied, the signal is correctly synchronized with the T0 detector. The distribution width is influenced by three main components: the STS time resolution, the T0 time resolution, and the spread in time-of-flight of particles from the event vertex to the STS sensors. The T0 has negligible resolution ($\sim$\qty{50}{ps}) compared to the STS. The contribution from the time-of-flight, given by the difference between the fastest and slowest particles in the event, was estimated with simulations and can reach up to \qty{2}{ns}. Therefore, the distribution width, represented by the error bar in the figure, indicates an upper limit of the STS time resolution.
        
        Figure~\ref{fig:time}, right part, shows the dependence of the distribution width on the signal amplitude. The distribution is wider for low-amplitude signals due to the walking effect and the lower signal-to-noise ratio. In contrast, once accounting for the time of flight, cleaner samples at larger signal amplitude allow the extraction of a time resolution around or better than \qty{5}{ns}.
    
    \subsection{Dead time} \label{ssec:dead_time}
        Different effects can be observed when two signals arrive close in time, depending on the time difference between them and the amplitude of the first signal. A pile-up in the first signal might occur for very close signals: the second signal is merged into the first, and its amplitude adds to the first signal. If the second signal arrives during dead time, it is lost. If the second signal arrives when the baseline is not restored, a new signal is created but with reduced amplitude.
        \begin{figure}
            \centering
            \includegraphics[width=0.6\textwidth]{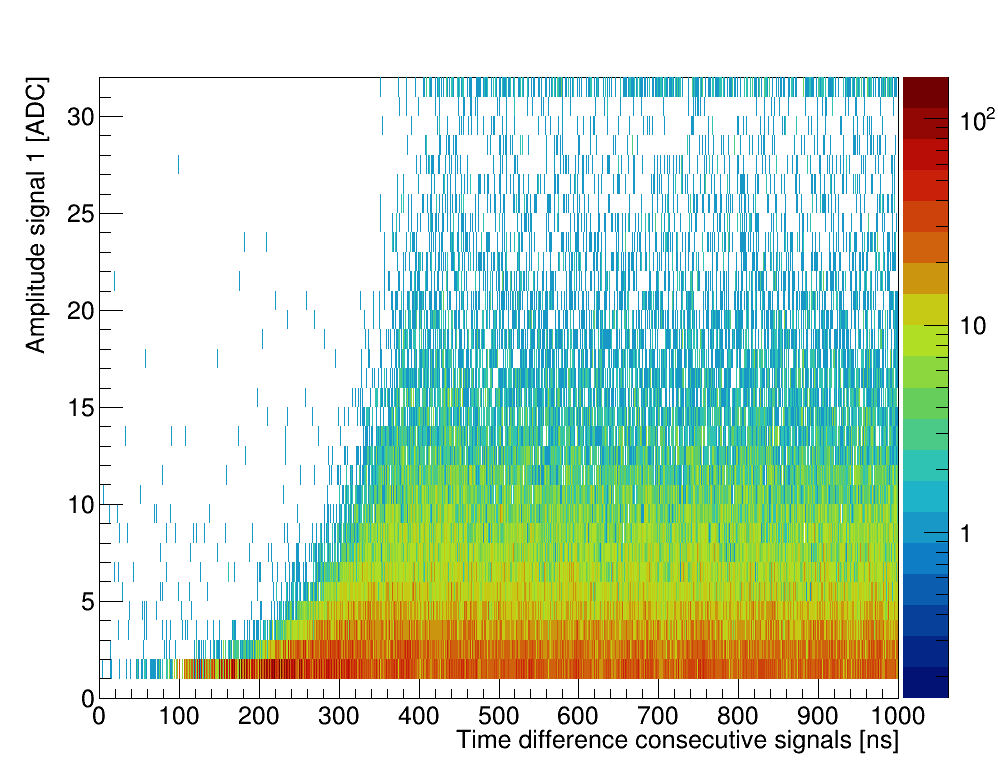}
            \caption{The amplitude of the first signal vs the time difference between consecutive signals in the same channel. The empty region corresponds to the detector dead time, which ranges from \qty{200}{ns} to \qty{350}{ns} for low to high charge values. Note that the numbering scheme of the STS XYTER discriminators (ADC) on the y-axis starts from 1.}
            \label{fig:dead_time}
        \end{figure}
        Figure~\ref{fig:dead_time} shows the time difference between two consecutive signals in the same channel versus the amplitude of the first signal. Operating the detector with the shortest shaping time of \qty{90}{ns} \cite{KASINSKI2018225}, a dead time between \qty{200}{ns} for low signal amplitude and \qty{350}{ns} is measured.
    
    \subsection{STS alignment} \label{ssec:alignment}
        Data-driven, software-based alignment of individual STS components provides sufficient precision to determine tracks and vertices, using only STS hits. This alignment is achieved through beam spot reconstruction as the figure of merit. As the mCBM setup does not include a magnetic field, any particle traverses the detector along a straight line. Every pair of hits in different stations forms a segment (tracklet) that can be extrapolated and projected to the target plane. The distribution of all tracklets extrapolations allows to reconstruct the beam spot. The vertex spread is defined as the RMS of the centers of such a distribution, found by each sensor pair. In the alignment procedure, independent sensor translations are applied through a gradient descent minimization, where the vertex spread is recalculated iteratively. The final translations, obtained at convergence after eight iterations, are consistent with the assembly precision of $\sim$\qty{100}{\um} and with the constraints from the mechanical units. After the alignment of all individual sensors, the overall primary vertex position, reconstructed via the beam spot method, is reported in Table \ref{tab:PV_resolution}.
    
    \subsection{Spatial Resolution} \label{space_resolution}
        The evaluation of the single-hit spatial resolution in a sensor requires measuring the differences between the reconstructed impact point by the device under test (DUT) and the predicted points obtained from track extrapolation. To evaluate spatial resolution, we exploit the spatial correlations between STS hits and utilize the outer detector systems (TRD and TOF) via a custom-modified 4D CA-tracking \cite{AkishinaTimeBasedRec2018} to reduce the number of false hit combinations. STS hits in the first and the last stations are used as reference. The station in the middle is used as the DUT. Using STS hits identified to belong to the same track allows for preserving the fake suppression that the tracking algorithm offers. Therefore, the evaluation of the track at the DUT is a simple straight-line extrapolation given by Eq.~\ref{eqt:x_trk}.

        \begin{align} \label{eqt:x_trk}
            \centering
            x_\text{DUT} &= x_i + T_x (z_\text{DUT} - z_i)\nonumber \\
            y_\text{DUT} &= y_i + T_y (z_\text{DUT} - y_i)
        \end{align}            

        where $T_x$, $T_y$ are the components of the track directional vector $\vec{T}$.        
        \begin{equation}
            \vec{T} = (T_x, T_y, 1) = (\frac{x_i - x_j}{z_i - z_j}, \frac{x_i - x_j}{z_i - z_j}, 1)
        \end{equation}        
        and $(x_i,y_i,z_i)$, $(x_j,y_j,z_j)$ are the space points given by the STS hits in the first and last station, respectively.
        
\begin{figure}[!h]
            \centering
            \includegraphics[width=0.47\textwidth]{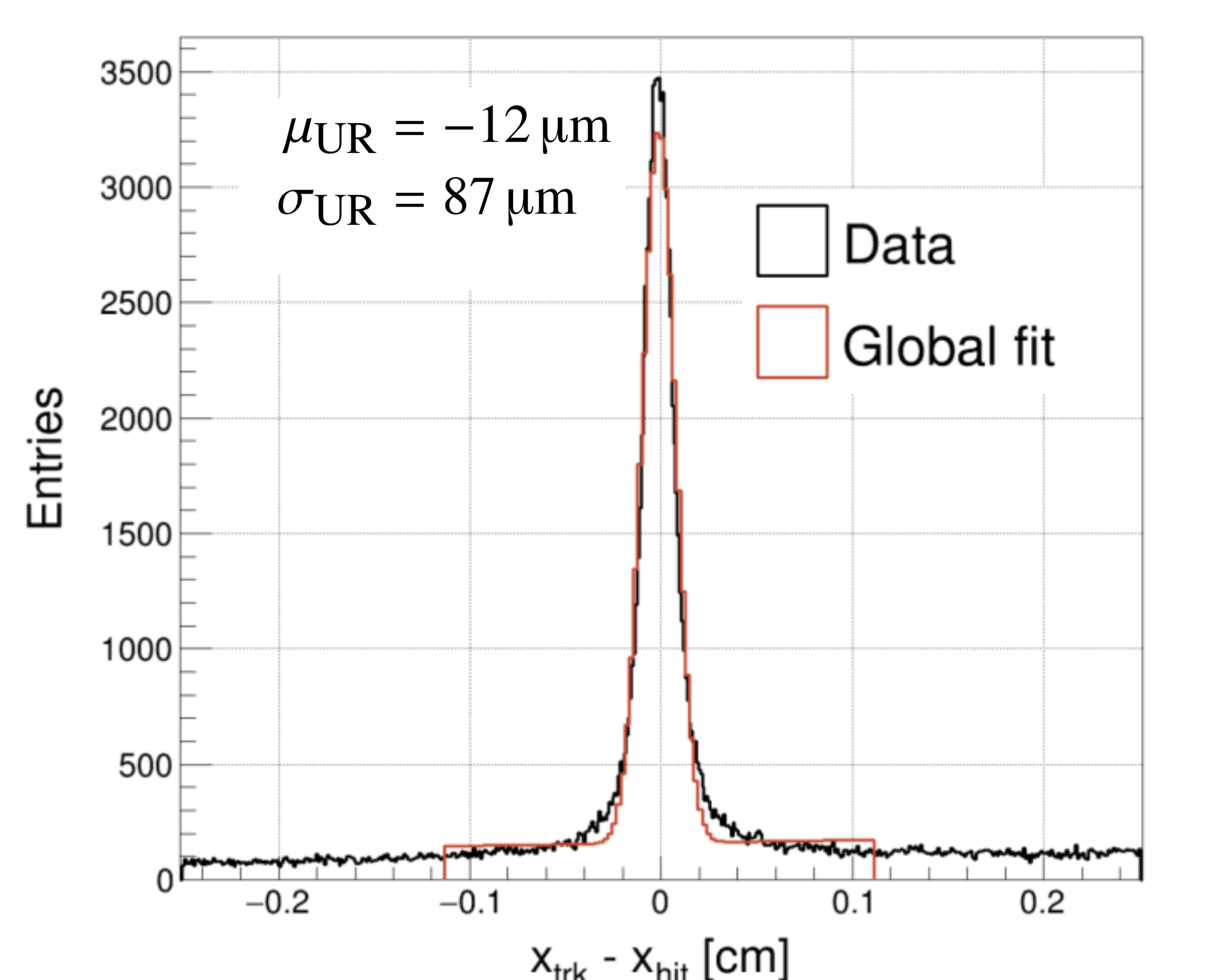}
            \includegraphics[width=0.47\textwidth]{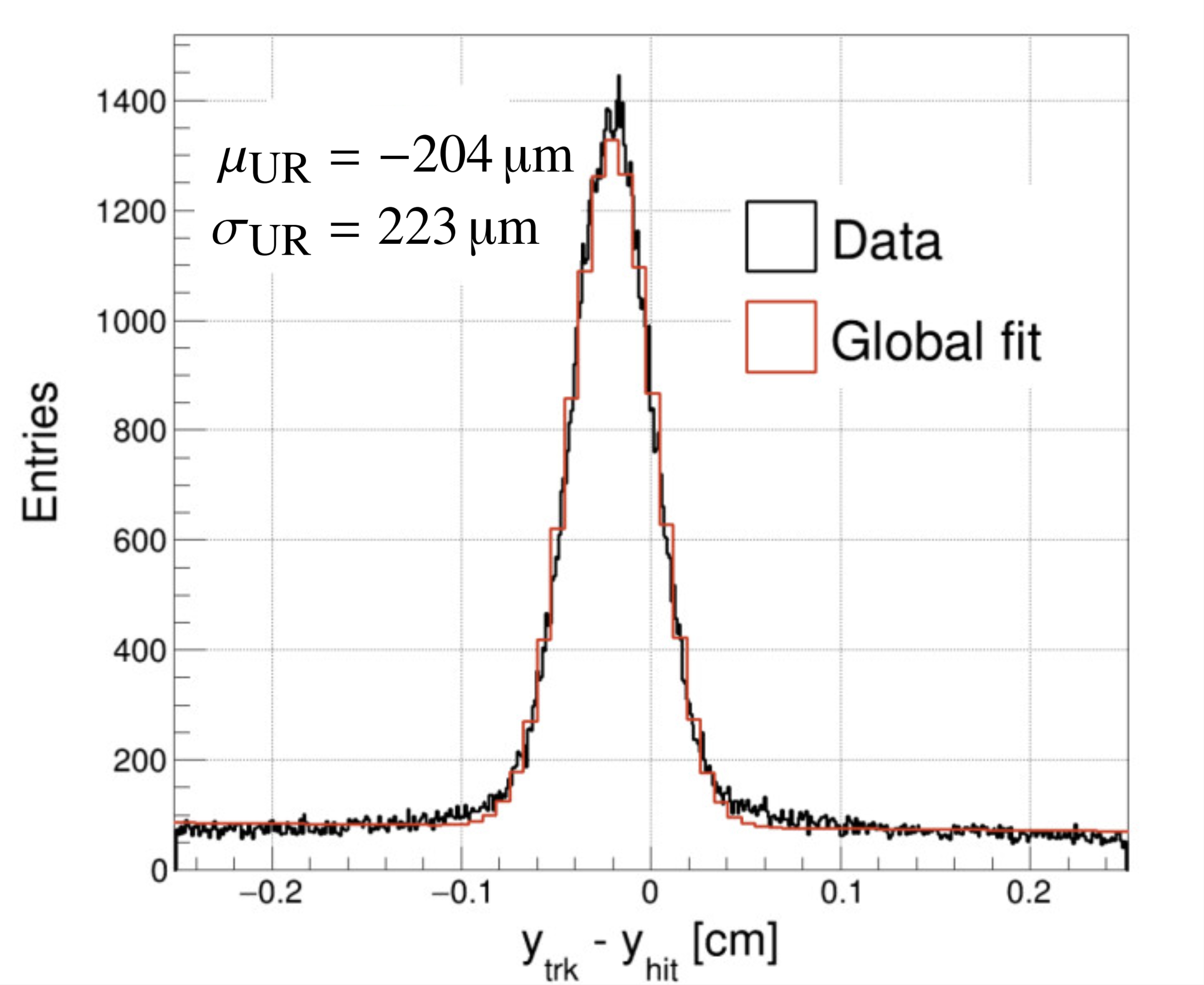}
            \caption{Unbiased hit-track residual distribution for a selected sensor module in X (left) and Y (right), respectively. The distribution is fitted by a Gaussian plus a second-order polynomial.}
            \label{fig:u1l0m0_res}
        \end{figure}

        The unbiased hit-track residual is the distance between the measured space point position on the DUT and the point obtained by extrapolating the track at the sensor plane. Figure~\ref{fig:u1l0m0_res} shows the unbiased hit-track residual distribution along the X (left) and Y (right) directions for a selected sensor. 
        
        In a system with ideal geometric alignment, the distribution spread $\sigma_\text{UR}$ is influenced by the resolution of the detector being tested $\sigma_\text{STS}$, the precision of track extrapolation $\sigma_\text{track}$, and the track deviation due to multiple Coulomb scattering $\sigma_\text{MS}$. 
        \begin{equation} \label{eq:sigmaUR}
            \sigma_\text{UR} = \sigma_\text{STS} \oplus \sigma_\text{track} \oplus \sigma_\text{MS} 
        \end{equation}
        
        The track extrapolation is done using only the first and last STS stations. Therefore, besides the location of the extrapolation plane, the accuracy $\sigma_\text{track}$ is influenced only by the spatial resolution of the STS modules ($\sigma_\text{STS}$) used to determine the track line.
        
        The impact of multiple Coulomb scattering on track residuals can be calculated with: 
        \begin{equation}
            \theta_0 = \frac{\qty{13}{\mathrm{MeV}}}{\beta\, c\, p} |z| \sqrt{\frac{x}{X_0}}\left[ 1 + 0.038 \ln{\left(\frac{x\, z^2}{X_0\, \beta^2} \right)}  \right]
        \end{equation}
        where  $p$, $\beta c$, and $z$ are the momentum, velocity, and charge number of the incident particle~\cite{Workman2022:Review}. The thickness of each module is such that $X / X_0$ = 0.38 $\pm$ 0.01\%. 
        
        The STS spatial resolution $\sigma_\text{STS}$ can then be extracted from the width of the residual distribution $\sigma_\text{UR}$ shown in Fig.~\ref{fig:u1l0m0_res}.
        A Gaussian fit added to a quadratic polynomial to describe the background provides a reasonably good characterization of the width of the central peak for both x and y distributions. The resolution obtained from the fit is given in Table~\ref{tab:sts_resolution}. However, the residual distribution deviates slightly from a Gaussian shape, even after accounting for a quadratic background. This deviation is mainly due to sensor misalignment and ghost hits. Two approaches have been implemented to address local dependencies and non-Gaussian effects. The first assumes that the true Gaussian distribution of the track-hit residuals dominates the peak, allowing extraction of $\sigma_\text{UR}$ from the FWHM relationship. The second approach calculates the average RMS of the residuals along the x and y coordinates of the sensor after subtracting the flat background. This method aims to prevent the broadening of the estimated $\sigma_\text{UR}$ by avoiding the superposition of multiple Gaussian peaks with different mean values, which results from residual misalignment. The resolution obtained with these other approaches, also given in Table~\ref{tab:sts_resolution}, is smaller and closer to the theoretical expectations of $58/\sqrt{12} = \qty{17}{\um}$.
        
        \begin{table}[]
            \caption{Spatial resolution of the STS $\sigma_\text{STS}$  in X and Y for different detectors of Station 1. Here, "L" refers to the ladder, and "M" to the module. The values are extracted from the width of the unbiased residuals, $\sigma_\text{UR}$ using Eq.~\ref{eq:sigmaUR} as: $\sigma$ from residual fit, FWHM/2.35, and average RMS. Reported values units are micrometers (\qty{}{\um}).}
            \begin{tabular}{c||c|c|c||c|c|c}
                sensor & $\sigma_{x}^{fit}$ & $\sigma_{x}^{FWHM}$ &$\sigma_{x}^{<RMS>}$ & $\sigma_{y}^{fit}$ &  $\sigma_{y}^{FWHM}$ &  $\sigma_{y}^{<RMS>}$ \\ \hline        
                L0M0 & 29 & 25 & 23 & 103 & 101 & 72 \\
                L0M1 & 28 & 27 & 26 & 102 & 105 & 69 \\
                L1M0 & 29 & 25 & 24 & 103 & 85  & 75 \\
                L1M1 & 28 & 29 & 27 & 102 & 110 & 72 \\
            \end{tabular}
            \label{tab:sts_resolution}
        \end{table}
    
    \subsection{Vertex reconstruction}\label{sec:vtx_rec}
        \begin{figure}[h!]
            \centering
            \includegraphics[width=.6\textwidth]{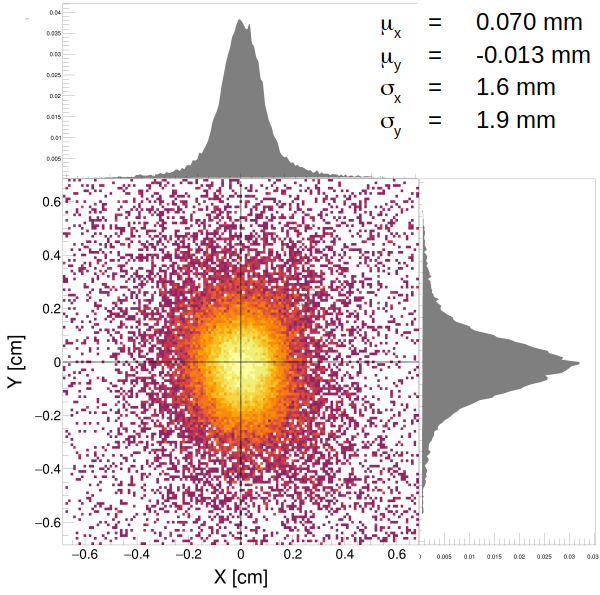}\quad 
            \caption{XY projection of the event primary vertex, together with one-dimensional distributions in X and Y, respectively. The mean and the widths of the distributions, obtained from a Gaussian fit, are displayed in the top right panel.}
            \label{fig:vertex_3D_xoy}
        \end{figure}
        
        In contrast to the beam spot, which is determined as an average over many events, the event-by-event primary vertex provides a precise estimate of the primary interaction point of a single collision. The three-dimensional primary vertex is reconstructed with the Point of Closest Approach (PCA) method, i.e., averaging the position of all valid PCAs of any track pairs in the event. A PCA is considered valid if the distance between the point and tracks is not much larger than the vertex resolution ($\sim100--\qty{500}{\um}$).
        
        \begin{figure}[htb]
                \centering
                \begin{subfigure}[b]{0.55\textwidth}
                    \centering
                    \includegraphics[width=.98\textwidth]{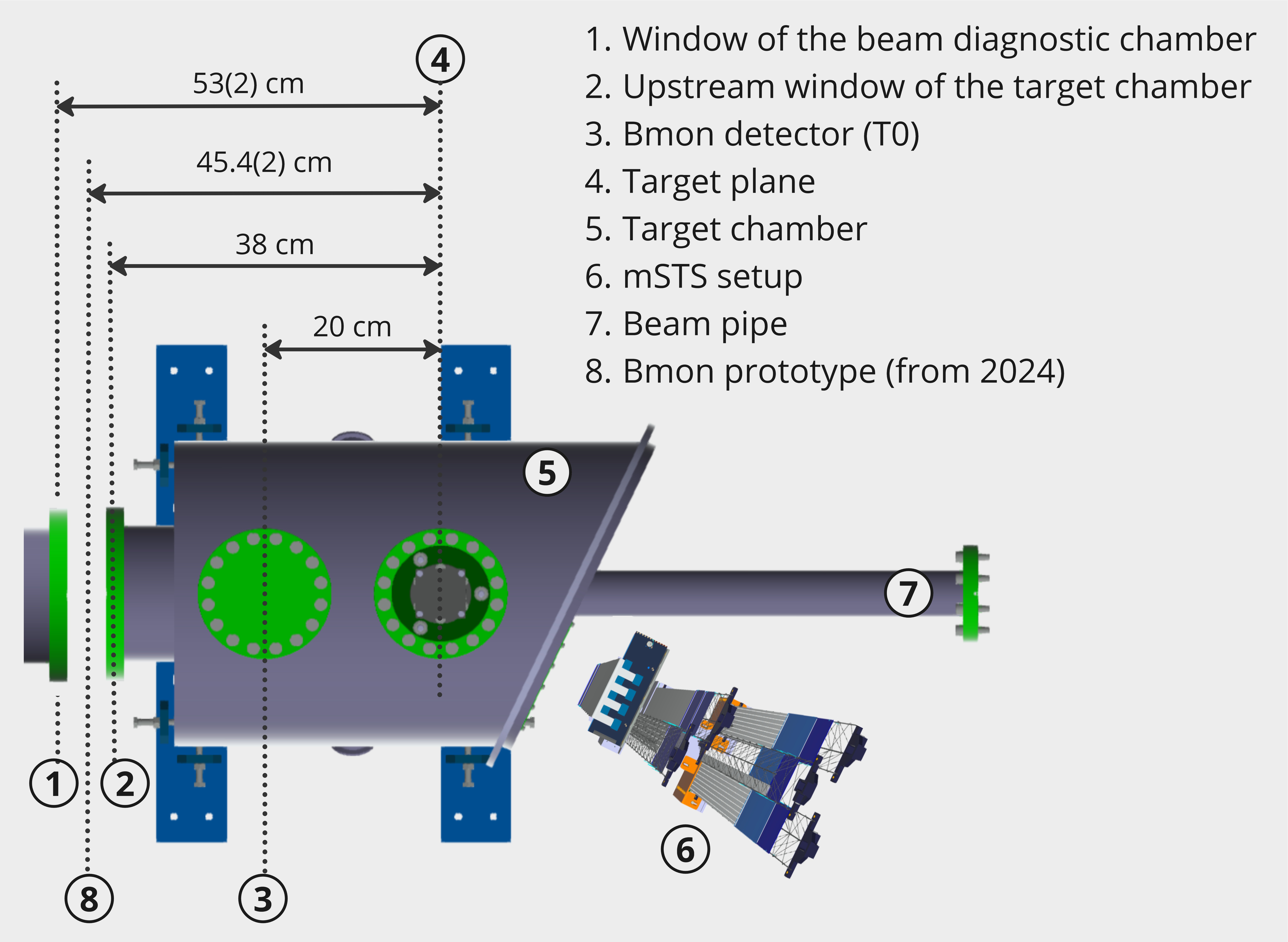}
                    \caption{}
                    \label{fig:mcbm_target}
                \end{subfigure}
                \begin{subfigure}[b]{0.43\textwidth}
                    \centering
                    \includegraphics[width=.98\textwidth]{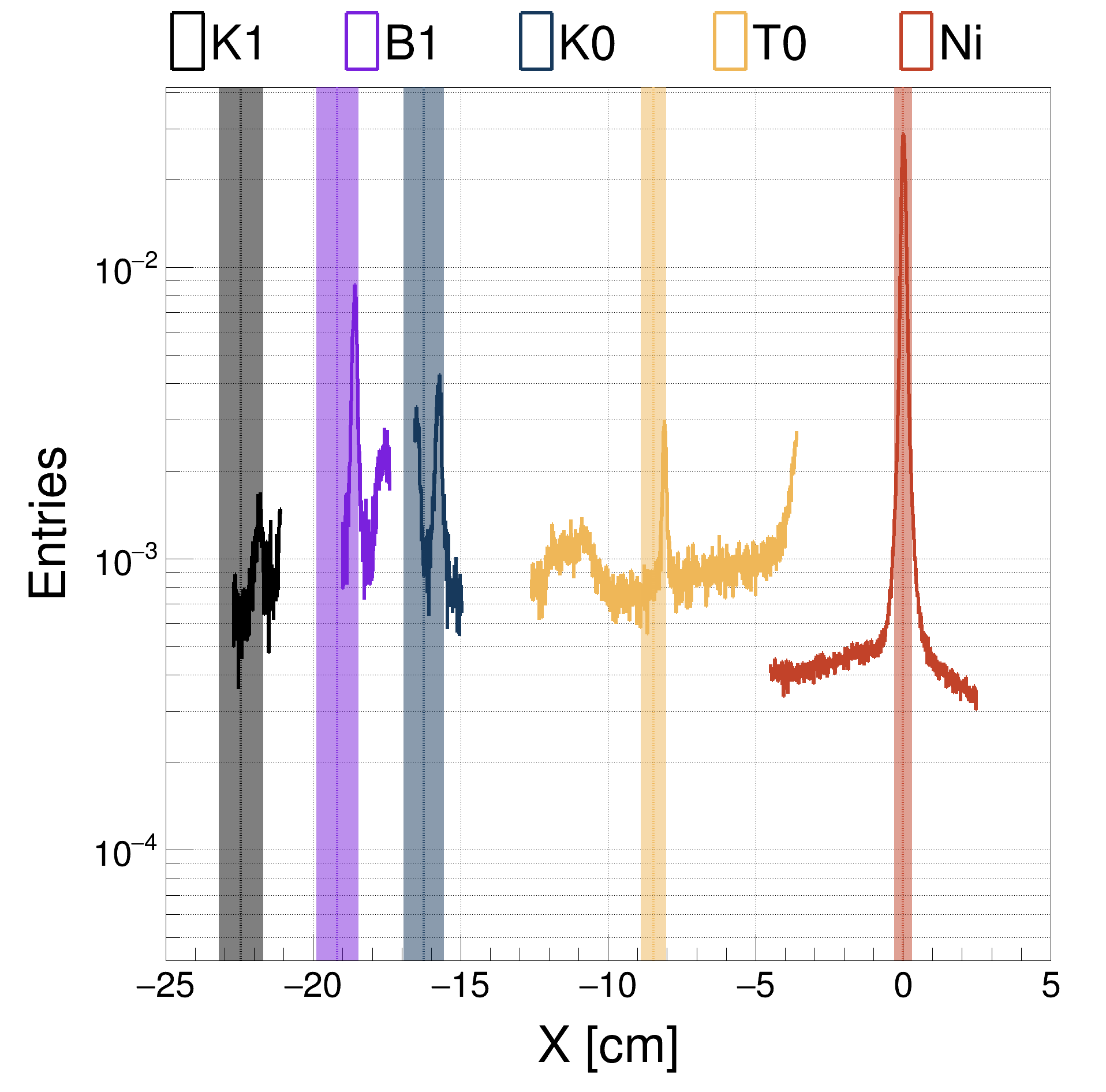}\quad
                    \caption{}
                    \label{fig:z_planes_peaks}
                \end{subfigure}                
                \caption{ (left) Zoom-in view of the target chamber region, highlighting the planes of interest for vertex reconstruction. (right) Projection of the reconstructed vertex along the x-axis for different z-planes. The dashed lines indicate the expected position of the vertex, together with the expected width, marked by the corresponding bands for the different secondary targets: Ni (the main Ni target), T0 (the T0 detector), B1 (the 2024 BMon prototype), K0 (the window of target chamber), K1 (the window of the diagnostic chamber).}
            \end{figure}

        Figure~\ref{fig:vertex_3D_xoy} shows the XY projection of the average position of PCA, together with the one-dimensional distributions in X and Y. The width of the primary vertex distribution, about \qty{1.5}{mm}, is dominated by the beam size.
        
        The primary vertex resolution is usually estimated by comparing the vertices reconstructed from a track sample that has been randomly split into two parts. Given the very low track multiplicity in the mCBM acceptance, we instead compare the vertex reconstructed using the PCA method to the mean of the beam spot reconstruction. As shown in Table \ref{tab:PV_resolution}, the results of vertex reconstruction using both methods differ only by \qty{50}{\um} in x and \qty{400}{\um} in y.
        
        \begin{table}[]
            \caption{Results of the vertex reconstruction with the Point-of-Closest-Approach (PCA) and the beam spot (BS) method. Reported values units are millimeters (\qty{}{mm}).}
            \begin{tabular}{c||c|c|c|c}
                method & $\mu_x$ & $\mu_y$ & $\sigma_x$ & $\sigma_y$ \\
                \hline
                PCA  & 0.070  & -0.013  & 1.6 & 1.9 \\
                BS   & 0.021 &  0.42 & 1.6 & 1.7\\
            \end{tabular}
            \label{tab:PV_resolution}
        \end{table}
                
        Taking advantage of the tracking algorithm and using tracks with two STS hits, the vertex distribution reveals distinct structures originating from secondary targets, corresponding to different interaction points within the target chamber, highlighted in Fig.~\ref{fig:mcbm_target}.

        The different structures can be focused on by changing the z position of the projection planes, depending on where the tracks originated. This can be done for the different secondary target planes (e.g., $\text{z}=\qty{-20}{cm}$ for the T0 detector).
        The projection along the x-axis of the vertex reconstruction for each plane is shown with different colors in Fig.~\ref{fig:z_planes_peaks}. A dashed line represents the expected position of such peaks, while the shaded box stems from the uncertainty in the projection of the beam spot to different z-planes, which increases with increasing distance in z.
        
        Besides minor deviations due to the detector's non-perfect alignment, the agreement between the measured vertices and the expected position is remarkably good.    
        \begin{figure}[!h]
            \centering
            \includegraphics[width=0.47\textwidth]{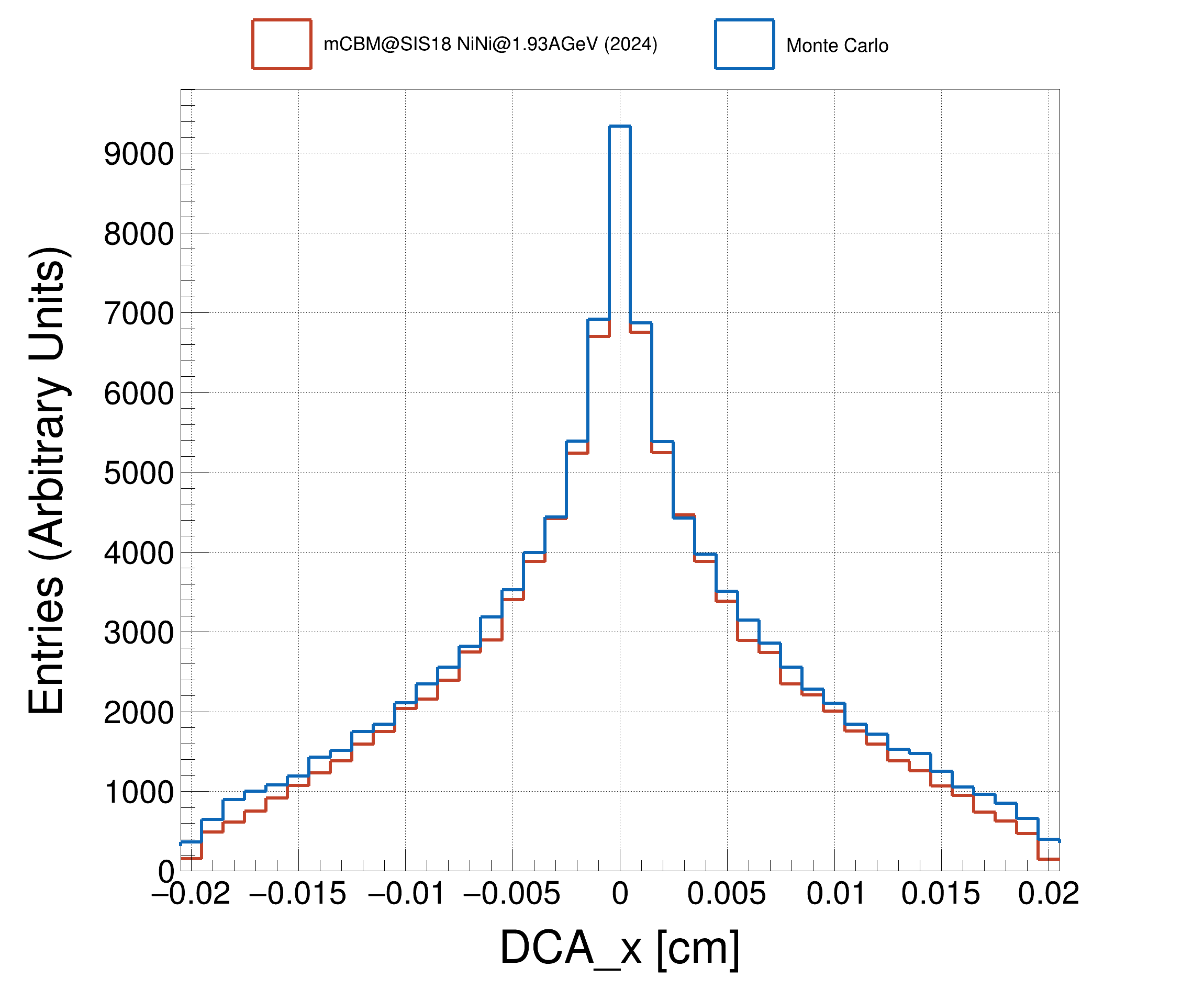}
            \includegraphics[width=0.47\textwidth]{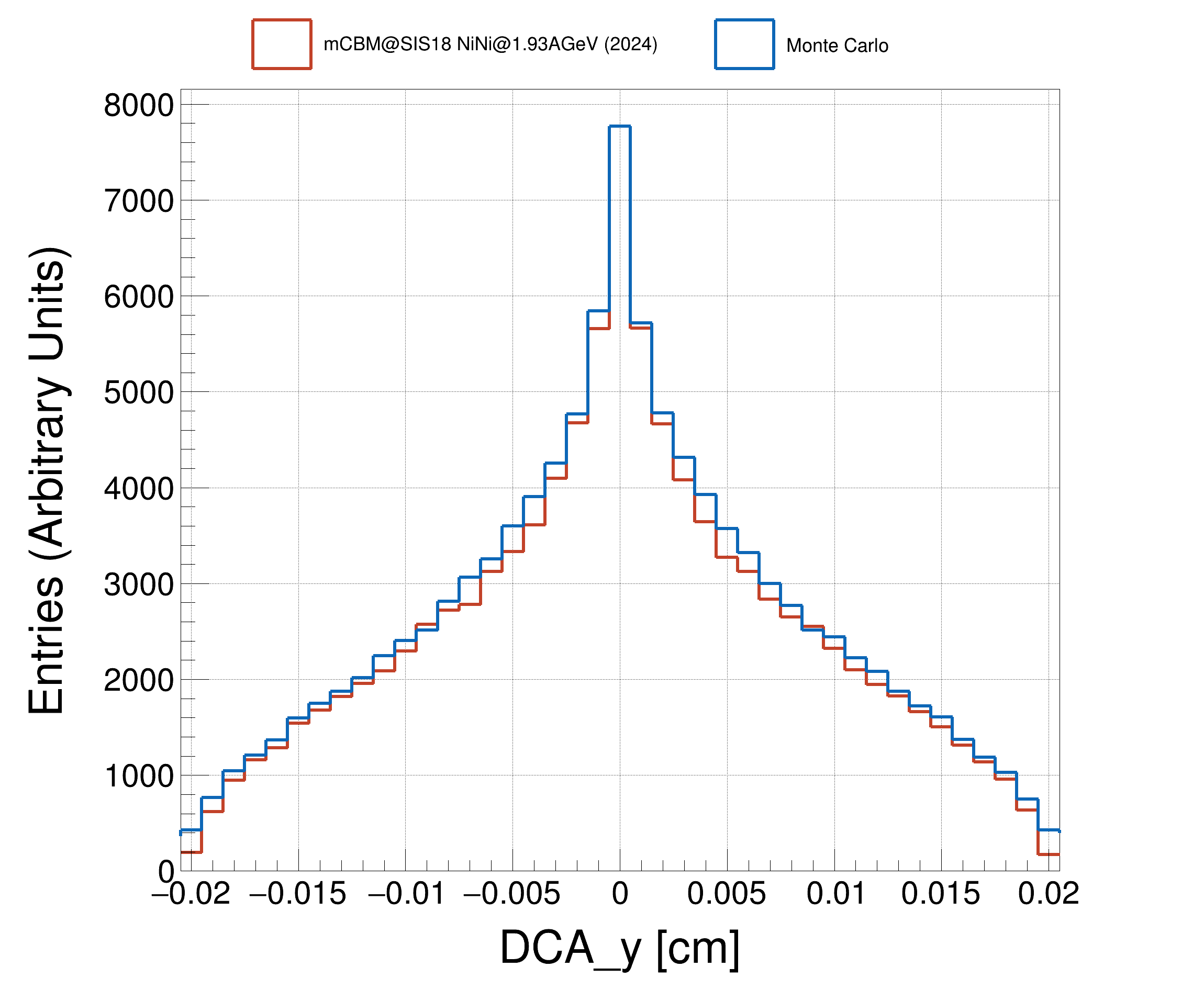}
            \caption{Track impact parameter distribution, in X (left) and Y (right), for real (red) and simulated (blue) data.}
            \label{fig:ImpactParameter}
        \end{figure}
        
    \subsection{Impact parameter} \label{ssec:dca}
        Track measurements with a precision of a few \qty{}{\um} near the interaction point allow the determination of decay vertices, which is crucial for reconstructing complex decay topologies.
        The impact parameter (IP) of a track is defined as the transverse distance of closest approach (DCA) of a particle trajectory to the primary vertex. Tracks from weak decays typically have larger IPs than those from the PV. Since there is no magnetic field in the mCBM setup, tracks are extrapolated to the primary vertex with a linear extrapolation. The IP resolution is governed by three main factors: multiple scattering of particles in detector material, the single hit resolution, and the distance between the PV and the closest measurement in the detector. The IP resolution can be determined by examining the widths of the distributions of the 1-D projections of the IP in x and y, shown in Fig.~\ref{fig:ImpactParameter}. The FWHM, proportional to uncertainty in the hit measurement, is about $\sim$\qty{50 (60)}{\um} in X (Y), respectively.
        
        The analysis is reproduced with Monte Carlo simulations, in which Ni+Ni collision events were generated using UrQMD \cite{BassUrQMD1998, BleicherUrQMD1999, PetersenUrQMD2008} and transported using GEANT3 \cite{Brun:1082634}. The primary vertex is smeared according to the distribution obtained from real data. The data were further digitized and processed with the same reconstruction chain as real data, Fig.~\ref{fig:vertex_3D_xoy}. The IP distributions, shown in Fig.~\ref{fig:ImpactParameter} for real data, are in excellent agreement with those extracted from MC simulations.
        
    \subsection{Hit reconstruction efficiency} \label{ssec:hit_rec_eff}
        The hit reconstruction efficiency corresponds to the probability of detecting a particle passing through the sensor. It can be determined as the measured fraction of hits on a track per possible hits along the track trajectory. In order to avoid bias from particles that decay (and therefore may have no signal in the first STS station) or are absorbed (and have no signal in the last STS station), this study refers to the middle STS station. Tracks used as references are those reconstructed with hits in all other detector layers, i.e., the first and last STS stations, and at least one hit in a TRD station and one in a TOF station to suppress fake combinations. Furthermore, tracks were requested to originate from the primary vertex, i.e., a cut of \qty{1.5}{mm} for the DCA of the track to the vertex and \qty{2.5}{mm} in the z direction. The efficiency is measured as a ratio of the number of tracks where all stations detected a hit over the number of tracks where at least the reference modules detected a hit.
        
        \begin{figure}
            \centering
            \includegraphics[width=0.6\textwidth]{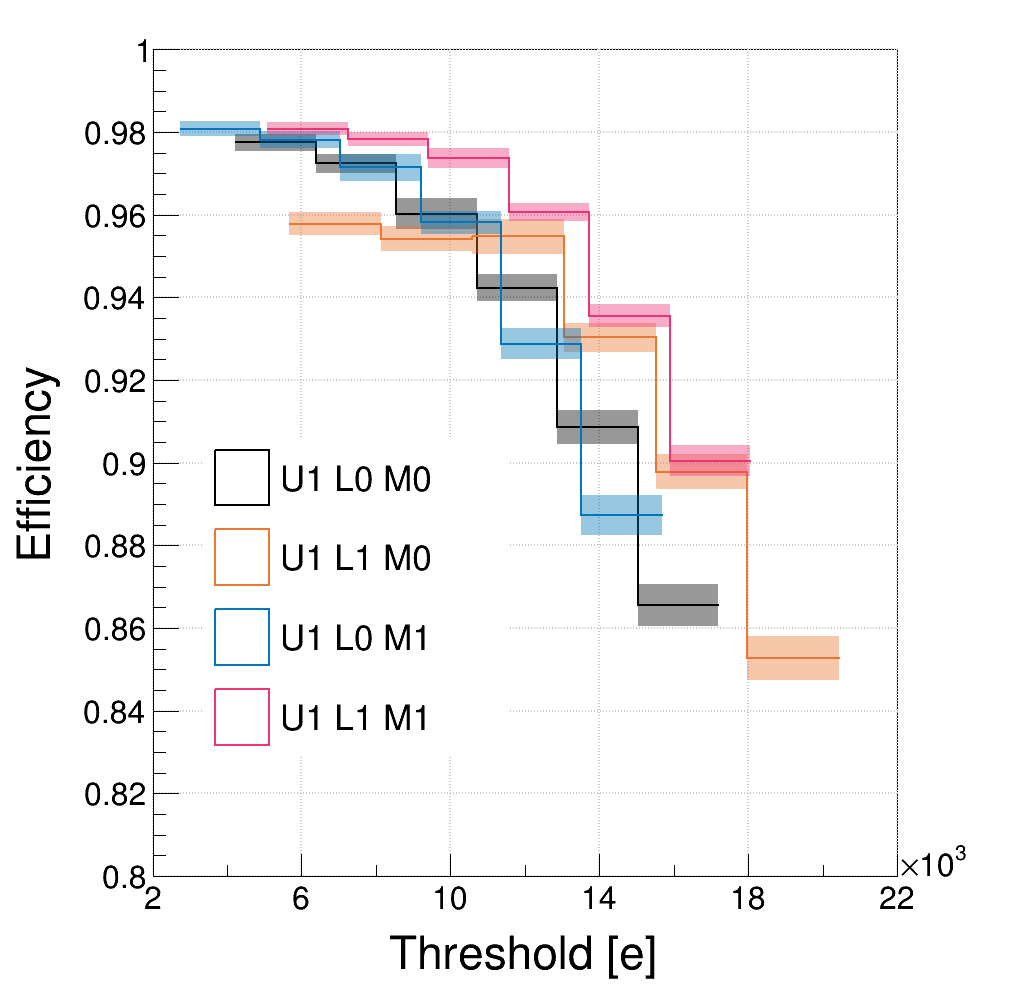}
            \caption{Hit reconstruction efficiency as a function of the threshold for different detectors of Station 1. 
            Here, "U1" refers to the Station, "L" refers to the ladder, and "M" refers to the module.
            The solid line represents the averaged efficiency for a well-selected active area within the sensor. The error band represents both statistical and systematic uncertainties.}
            \label{fig:hre_vs_thr}
        \end{figure}
        
        The tracks found using only the reference stations were interpolated to the test module. Only tracks with the projected impact point within the test sensor’s sensitive area are used. The region around the projected impact point was searched for hits using the fit results of the hit-track residual distribution. A hit is considered to belong to a track if its position is within \qty{3}{\sigma} from the extrapolated impact point, as from Fig.~\ref{fig:u1l0m0_res}. 
        
        The efficiency was found to be very uniform across the sensor active area, while the missing hits are concentrated along noisy or dead strips of the module. The particle loss is thus a direct consequence of the module production defects. The efficiency of the module areas with nominal strip noise and absence of bonding defects is about 98\%.
        
        \begin{figure}[!h]
            \centering
            \includegraphics[width=0.5\textwidth]{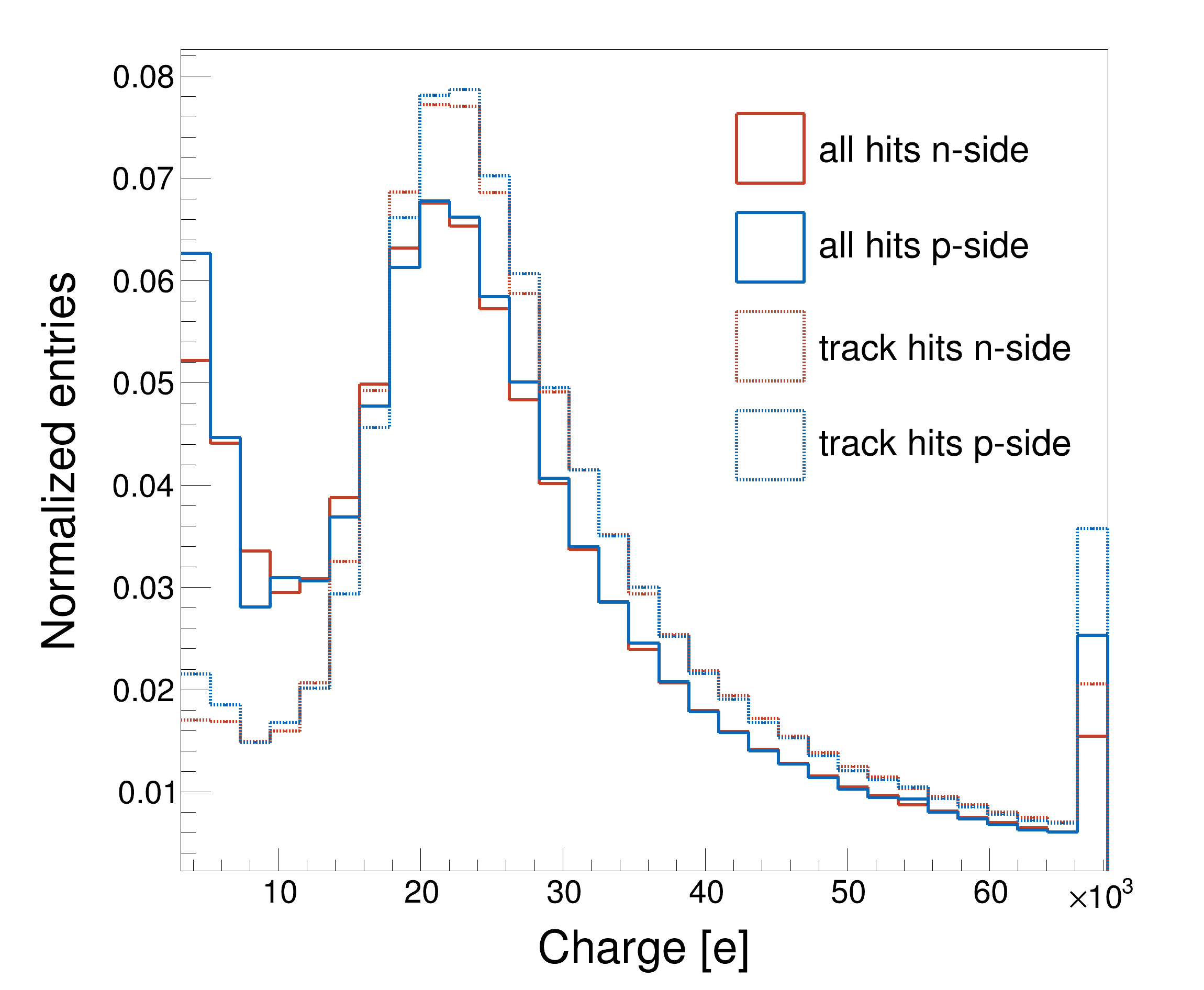}
            \includegraphics[width=0.4\textwidth]{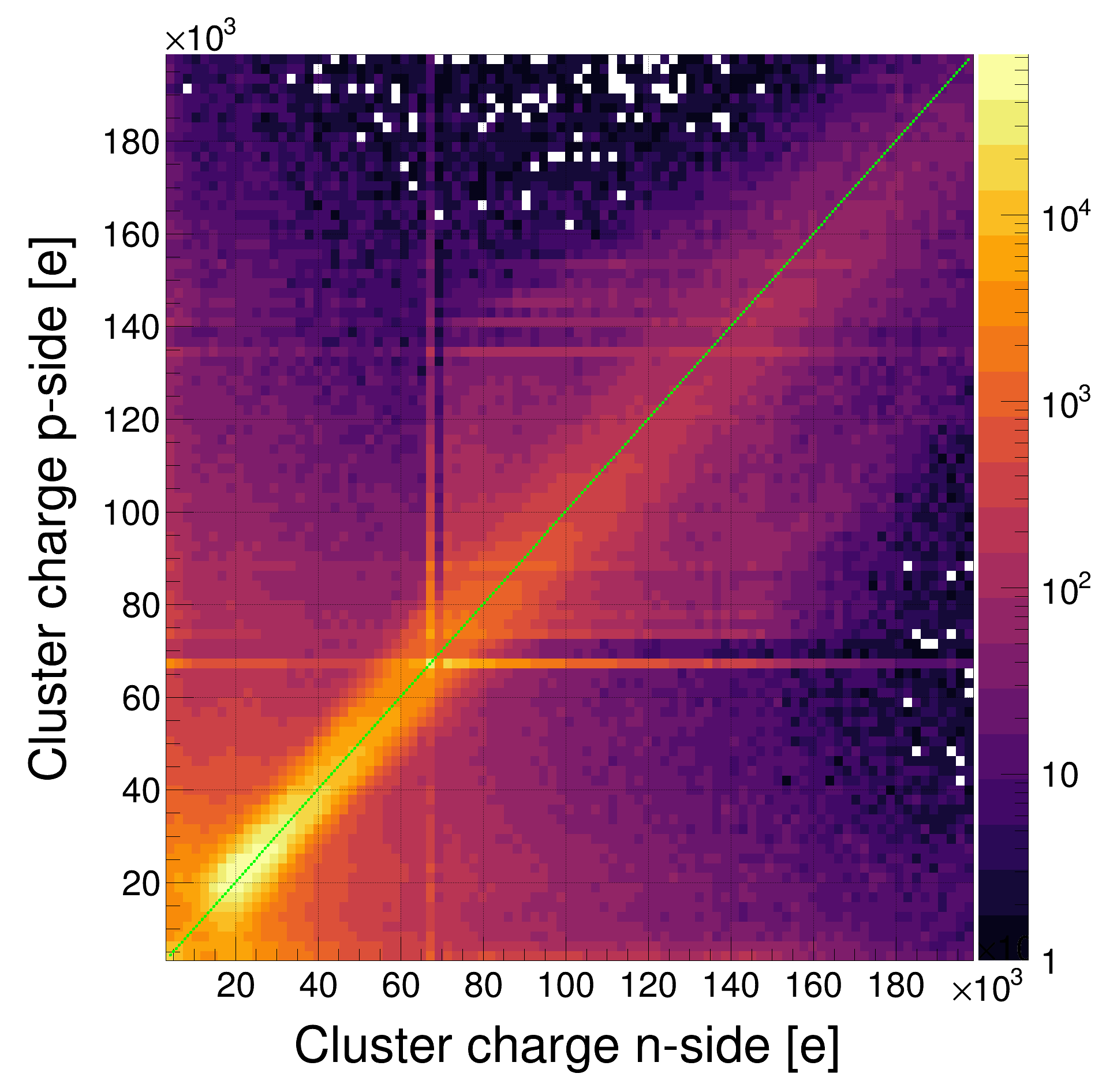}
            \caption{Top: Signal amplitude distribution for n- and p-side. The overflow bin is filled at the upper end of the charge digitization range. Bottom: correlation between the signal amplitude on both sides of a module. Clusters that include one or more signals in the overflow bin result in vertical and horizontal lines at multiples of the digitization range.}
            \label{fig:charge}
        \end{figure}
        
        Furthermore, the threshold dependence of the hit reconstruction efficiency was studied. 
        If the signal amplitude deposited in a given strip is below the digitization threshold, the signal in that channel is not recorded, resulting in a loss of information. If a major part of the energy deposited by the particle's passage through the detector is recorded by neighboring strips, only the spatial resolution is worsened; however, increasingly high thresholds will result in a reduction of hit detection efficiency.
        For these studies, the entire data analysis was performed for discrete increments of the digitization thresholds of raw strip signals. Figure~\ref{fig:hre_vs_thr} shows the hit reconstruction efficiency for the modules in the middle station as a function of the digitization threshold. 

    \subsection{Amplitude measurement and signal-to-noise ratio} \label{ssec:noise}
        The measured signal amplitude is compared to the noise measurements determined beforehand in the module characterization to evaluate the signal-to-noise ratio.
        The noise (N) is quantified as the average equivalent noise charge of the modules from the smearing of the discriminator response to a given pulse value. It amounts to around \qty{1000}{e} ENC, in agreement with the estimation based on the total detector capacitance \cite{RodriguezRodriguez2024}. The signal (S) is determined by the most probable charge amplitude produced by minimum ionizing particles produced in the collision.
        
        Figure~\ref{fig:charge} shows the signal amplitude distribution for p- and n-side. The left panel shows the correlation between the signal amplitude on both sides of the sensor. The readout ASICs of the STS modules were calibrated in a charge range of approximately \qty{10}{fC} (\qty{65}{ke}), with ADC gain of 0.335$\pm$\qty{0.003}{fC/LSB} (about \qty{2000}{e/LSB}) both for n- and p-side. The modules were operated at rather low thresholds (down to \qty{3}{\sigma}, i.e., approximately \qty{3000}{e}).
        The signal amplitude is shown for all reconstructed hits and, in order to minimize noise and background, only for those hits that were further correlated by tracking to outer detectors. This signal corresponds to the typical admixture produced in heavy-ion collisions of pions, protons, and kaons. The remaining contribution from noise is visible in the lower part of the distribution, i.e., below \qty{10}{ke}, largely suppressed when requiring that the hits belong to fully reconstructed tracks.
        The average signal (S) is extracted by fitting the distributions with a Landau function, resulting in a most probable value of 22.7$\pm$\qty{1.4}{ke}. For the estimation of the errors, the ADC gain, threshold spread, and the systematic error of the fitting were taken into account. The statistical error of the fitting was considered negligible with respect to the systematic error.
        Using the corresponding noise measurements of about \qty{1000}{e} ENC, a signal-to-noise ratio S/N around 23 is determined.
                
\section{Summary} \label{sec:summary}
    The operational performance of the prototype STS system for the CBM experiment was studied in a heavy-ion collision experiment with a beam energy of around 1--\qty{2}{AGeV} at the mCBM setup at SIS18. Precision track and vertex reconstruction has been demonstrated in the hit-track residual distribution, vertex position, and transverse impact parameter resolutions. Given some residual misalignment in these prototype experiments, a spatial resolution around \qty{25}{\um} is consistent with the theoretical estimate of the coordinate resolution of the sensor with \qty{58}{\um} strip pitch. A timing resolution around or better than \qty{5}{ns} has been demonstrated. The deadtime was estimated between \qty{200}{ns} and \qty{350}{ns}. The readout electronics of the modules demonstrated stable operation under conditions of a detector load of \qty{56}{MHz/s/channel}. Operating the detector with low thresholds, around 3--\qty{4}{\sigma} RMS amplitude of the noise, the typical dark rate observed was around \qty{0.5}{kHit/s/channel}, which is less than 1\% of the maximum measured or the expected bandwidth saturation limit. In these conditions, the efficiency of hit reconstruction is found to be 98\% for the areas where all channels were operational. A signal-to-noise ratio above 20 was measured.
    
    In conclusion, the operational performance of the Silicon Tracking System detector, as evaluated through prototype experiments, aligns with the requirements set for the upcoming CBM experiment. The detector meets the demanding standards for precise tracking, vertex determination, momentum reconstruction, and data throughput, ensuring its readiness to support the CBM experiment's experimental needs.


\printcredits


\bibliographystyle{ieeetr} 

\bibliography{main}

\begin{thebibliography}{10}

\bibitem{CbmPhyProg2017}
T.~Ablyazimov and et~al., ``Challenges in {QCD} matter physics - the scientific
  programme of the compressed baryonic matter experiment at fair,'' {\em The
  European Physical Journal A}, vol.~53, p.~60, 2017.

\bibitem{StsTDR2013}
J.~Heuser, W.~Müller, V.~Pugatch, P.~Senger, C.~J. Schmidt, C.~Sturm, and
  U.~Frankenfeld, eds., {\em [{GSI} Report 2013-4] Technical Design Report for
  the {CBM} {S}ilicon {T}racking {S}ystem ({STS})}.
\newblock Darmstadt: GSI, 2013.

\bibitem{RodriguezRodriguez2024}
A.~R. Rodríguez, O.~M. Rodríguez, J.~Lehnert, A.~Toia, M.~Teklishyn,
  A.~Lymanets, D.~R. Garcés, J.~M. Heuser, and C.~J. Schmidt, ``Functional
  characterization of modules for the silicon tracking system of the {CBM}
  experiment,'' {\em Nuclear Instruments and Methods in Physics Research
  Section A: Accelerators, Spectrometers, Detectors and Associated Equipment},
  vol.~1058, p.~168813, 2024.

\bibitem{HamamatsuPhotonics}
``Hamamatsu photonics k.k.,'' 2023.

\bibitem{KASINSKI2018225}
K.~Kasinski, A.~Rodriguez-Rodriguez, J.~Lehnert, W.~Zubrzycka, R.~Szczygiel,
  P.~Otfinowski, R.~Kleczek, and C.~Schmidt, ``{C}haracterization of the
  {STS/MUCH-XYTER2}, a 128-channel time and amplitude measurement {IC} for gas
  and silicon microstrip sensors,'' {\em Nuclear Instruments and Methods in
  Physics Research Section A: Accelerators, Spectrometers, Detectors and
  Associated Equipment}, vol.~908, pp.~225--235, 2018.

\bibitem{mCBM_SIS18_Proposal}
{CBM Collaboration}, ``{mCBM@SIS18},'' {\em Darmstadt: CBM Collaboration},
  p.~GSI 58 S. (2017)., 2017.

\bibitem{TrdTDR2018}
{CBM Collaboration}, ``The transition radiation detector of the {CBM}
  {E}xperiment at {FAIR} : {T}echnical {D}esign {R}eport for the {CBM}
  {T}ransition {R}adiation {D}etector ({TRD}),'' Tech. Rep. FAIR Technical
  Design Report, Darmstadt, 2018.

\bibitem{TofTDR2014}
N.~Herrmann, ed., {\em {T}echnical {D}esign {R}eport for the {CBM}
  {T}ime-of-{F}light {S}ystem ({TOF})}.
\newblock Darmstadt: GSI, 2014.

\bibitem{RichTDR2013}
``Technical design report for the {CBM} ring {I}maging {C}herenkov
  {D}etector,'' tech. rep., 2013.

\bibitem{CuvelandDAQ2022}
``{T}echnical {D}esign {R}eport for the {CBM} {O}nline {S}ystems – {P}art
  {I}, {DAQ} and {FLES} {E}ntry {S}tage,'' PUB:(DE-HGF)3 / PUB:(DE-HGF)29~-,
  Darmstadt, 2023.

\bibitem{FrieseClusterFinding2019}
V.~Friese, ``A cluster-finding algorithm for free-streaming data,'' {\em EPJ
  Web of Conferences}, vol.~214, p.~01008, 9 2019.

\bibitem{Malygina_2017}
H.~Malygina and V.~Friese, ``A precision device needs precise simulation:
  Software description of the {CBM} silicon tracking system,'' {\em Journal of
  Physics: Conference Series}, vol.~898, p.~042022, 10 2017.

\bibitem{AkishinaTimeBasedRec2018}
V.~Akishina, I.~Kisel, I.~Vassiliev, and M.~Zyzak, ``Time-based reconstruction
  of free-streaming data in cbm,'' {\em EPJ Web of Conferences}, vol.~173,
  p.~04002, 2 2018.

\bibitem{Workman2022:Review}
R.~L. Workman and et~al., ``Review of particle physics,'' {\em PTEP},
  vol.~2022, p.~083C01, 2022.

\bibitem{BassUrQMD1998}
A.~Bass and et~al., ``Microscopic models for ultrarelativistic heavy ion
  collisions,'' {\em Prog. Part. Nucl. Phys.}, vol.~41, pp.~225--370, 1998.

\bibitem{BleicherUrQMD1999}
M.~Bleicher and et~al., ``Relativistic hadron-hadron collisions in the
  ultra-relativistic quantum molecular dynamics model,'' {\em J. Phys. G: Nucl.
  Part. Phys.}, vol.~25, pp.~1859--1896, 1999.

\bibitem{PetersenUrQMD2008}
H.~Petersen, J.~Steinheimer, G.~Burau, M.~Bleicher, and H.~Stöcker, ``Fully
  integrated transport approach to heavy ion reactions with an intermediate
  hydrodynamic stage,'' {\em Phys. Rev. C}, vol.~78, p.~044901, 2008.

\bibitem{Brun:1082634}
R.~Brun, F.~Bruyant, F.~Carminati, S.~Giani, M.~Maire, A.~McPherson,
  G.~Patrick, and L.~Urban, {\em {GEANT}: {D}etector {D}escription and
  {S}imulation {T}ool}.
\newblock CERN Program Library, Geneva: CERN, 1993.
\newblock Long Writeup W5013.

\end{thebibliography}



\end{document}